\documentclass[conference]{IEEEtran}
\usepackage{graphicx}
\usepackage{amsmath, cite}

\DeclareMathOperator{\sgn}{sgn}

\def\comment#1{}

\def\C(#1){|#1|}

\def\comment#1{\mbox{\;\;\;#1}}
\def\Ind(#1){\Delta\left[#1\right]}

\def\Inner(#1,#2){\langle #1, #2 \rangle}

\begin{document}
\title{BICM Performance Improvement via Online LLR Optimization}

\author{\authorblockN{Jinhong Wu, Mostafa El-Khamy, Jungwon Lee and Inyup Kang}
\authorblockA{Samsung Mobile Solutions Lab\\
San Diego, USA 92121\\
Email: $\{$Jinhong.W, Mostafa.E, Jungwon2.Lee,
Inyup.Kang$\}$@samsung.com}}


\maketitle

\begin{abstract}

We consider bit interleaved coded modulation (BICM) receiver performance improvement based on the concept of generalized mutual information (GMI). Increasing achievable rates of BICM receiver with GMI maximization by proper scaling of the log likelihood ratio (LLR) is investigated. While it has been shown in the literature that look-up table based LLR scaling functions matched to each specific transmission scenario may provide close to optimal solutions, this method is difficult to adapt to time-varying channel conditions. To solve this problem, an online adaptive scaling factor searching algorithm is developed. Uniform scaling factors are applied to LLRs from different bit channels of each data frame by maximizing an approximate GMI that characterizes the transmission conditions of current data frame. Numerical analysis on effective achievable rates as well as link level simulation of realistic mobile transmission scenarios indicate that the proposed method is simple yet effective.


\end{abstract}

\section{Introduction}

 	Bit interleaved coded modulation (BICM) has been widely adopted in wireless standards, e.g. $3$GPP High Speed Packet Access (HSPA) and Long-Term Evolution (LTE) \cite{3G2585,3G36212}. For multi-level modulation schemes, BICM transmission can be decomposed into multiple parallel bit channels by which each bit channel is capable of transmitting at certain rates according to the bit channel's transition probability density function. Achievable rates of BICM can be characterized by the generalized mutual information (GMI) \cite{TIT09_Martinez_Fabregas_Caire_Willems}, which is equivalent to the sum of GMI of individual bit channels. With the assumptions of perfect channel state information and optimum detection, BICM capacity has been shown to be achievable \cite{TIT09_Martinez_Fabregas_Caire_Willems} \cite{TCOM11_Nguyen_Lampe}.

 In real systems, however, with imperfections in the receiver, e.g., limited depth interleaving, imperfect channel estimation, sub-optimal detection, etc., the actual achievable rate may be substantially lower than the BICM capacity. Such characteristics are categorized as \emph{BICM with mismatched decoding} \cite{TIT09_Martinez_Fabregas_Caire_Willems}. On the other hand, by certain LLR correction prior to decoding, the achievable rate can be improved \cite{TCOM11_Nguyen_Lampe} \cite{ITW10_Jalden_Fertl_Matz}. As shown in \cite{ITW10_Jalden_Fertl_Matz} \cite{TCOM11_Nguyen_Lampe}, optimal correction is to obtain LLRs that are consistent with the overall bit-channel transition probability between each coded bit and the detector output LLR. It is suggested in above references that LLR correction may be performed by off-line data analysis to first find empirical probability density function (PDF) of LLRs from each bit-channel, then numerically scale the LLRs. Specifically, a look-up table (LUT) with scaling factors matched to LLRs from each bit channel can be generated. For simpler processing, uniform scaling factors may be used instead of LUTs. Generally, such off-line analysis based scaling methods are valid to scenarios where transmission conditions are known to the receiver. In typical mobile communication systems, however, accurate identification of transmission scenarios in real-time is difficult due to rapid changes in channel conditions. As a result, such methods may not be applicable, as applying mismatched scaling factors to LLRs can degrade decoder performance.

 In this paper, we consider an adaptive method for linear LLR optimization of a mismatched BICM receiver. We follow the GMI concept and seek to linearly optimize LLRs by scaling factors obtained from simple online numerical search. In this process, decoder decision feedback is used to calculate an approximate GMI to search for scaling factors. As will be shown, this online search method provides uniform scaling factor(s) per bit channel that can adapt to time-varying transmission scenario without explicitly knowing the channel conditions.

 The paper is organized as follows. Section-\ref{sec.sysmodel} introduces the BICM system model and discusses achievable rates by practical receivers. Section-\ref{sec:off_line} reviews and discusses existing scaling methods based on off-line channel analysis. Section-\ref{sec:GMI_scaling} introduces the proposed scaling algorithm. Section-\ref{sec:simu} presents simulation results. Finally, Section-\ref{sec:conclusion} concludes the paper.

\section{BICM System model and achievable rates}\label{sec.sysmodel}

\begin{figure}[h!]
\centering
\includegraphics[width=3.5in]{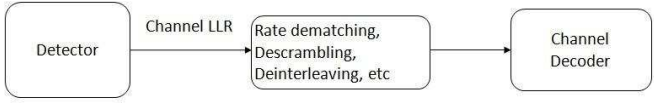}
\caption{A simplified block diagram of BICM receiver}
\label{fig:BICM_Generic}
\end{figure}

We consider bit-interleaved channel-coded modulation transmitted over memoryless channel. Information bit sequence $\mathbf{u}=[u_{0},u_{1},\cdots,u_{K-1}]$ is encoded to produce coded bit sequence $\mathbf{c}=[c_{0},c_{1},\cdots,c_{K/R-1}]$, where $R$ is code rate. Proper interleaving and possibly scrambling of $\mathbf{c}$ produces a new bit sequence $\mathbf{b}=[b_{0},b_{1},\cdots,b_{K/R-1}]$, which is then mapped into symbols from a $M$-ary signal constellation $\chi$, producing $\mathbf{x}=[x_{0},x_{1},\cdots,x_{N-1}]$, where $x_{n}\in\chi$. For simplicity, we consider quadrature amplitude modulation (QAM), but extension to other modulation schemes is straightforward. The modulated signals are transmitted over the wireless channel via single or multiple antennas after layer mapping and precoding. At the receiver side, following detection of received signal, $\mathbf{y}=[y_{0},y_{1},\cdots,y_{N-1}]$, a general process of descrambling, rate de-matching, and de-interleaving recovers the LLR sequence to match that of the coded bits. The channel decoder then takes these LLRs as input. A generic description of such a process is shown in Fig.~\ref{fig:BICM_Generic}. 

The total BICM channel may be decomposed into $m=\log_{2}M$ parallel binary-input channels. Note that for square QAM with Gray mapping, bit loading in the real and imaginary dimensions of the complex signal are identical. Therefore, we only need to consider $m/2$ distinct bit channels. 


\subsection{Achievable rates of BICM with matched and mismatched decoding metrics}\label{subsec:achievableRates}

The BICM capacity is given by \cite{TIT09_Martinez_Fabregas_Caire_Willems}

\begin{equation}\label{BICM_capacity}
\begin{split}
    C^{BICM}&= \sum\limits^{m}_{j=1}I(B_{j};Y)\\
    &=-\sum\limits^{m-1}_{i=0}E_{X,Y}\left\{\log\sum\limits_{b\in\mathcal{B}}p_{B_{i}}(b)\left(\frac{p_{Y|B_{i}}(Y|b)}{p_{Y|B_{i}}(Y|b_{i}(X))}\right)\right\}
\end{split}
\end{equation}
where $I(B_{j};Y)$ is the mutual information between the random variable of transmitted bit through $j$-th bit channel, $B_{j}$, and the received signal $Y$, $p_{B_{i}}(b)$ is the probability of bit $b$ transmitted through the $i$-th bit channel, and $p_{Y|B_{i}}(Y|b)$ is the conditional PDF of receiving $Y$ given $b$ is transmitted. The above rate is achievable from the argument of random coding with typical sequences. Note that \eqref{BICM_capacity} indicates that the capacity of BICM systems is the sum of capacities of the modeled individual bit channels.
%
With perfect interleaving and optimal detection, the detector output LLRs are sufficient statistics of transition probabilities $p_{Y|X}(y|x)$. For optimal detection, the definition of LLR is

\begin{equation}\label{LLR_def}
    L^{opt}_{B_{i},Y} = \ln\frac{p_{B_{i}|Y}(b=1|y)}{p_{B_{i}|Y}(b=0|y)}.
\end{equation}


In real systems, however, decoding metrics in form of detection output LLRs, are generally mismatched to the true channel transition probabilities $p_{Y|B_{i}}(y|b_{i})$. Another decoding metric is introduced that works as an estimate of the bit channel transition probability
\begin{equation}\label{mis_prob_LLR}
\begin{split}
   q_{B_{i},Y}(b_{i}(x),y) &= \hat{p}_{Y|B_{i}}(y|b_{i})
\end{split}
\end{equation}
and the available LLR is therefore an expression of
\begin{equation}\label{LLR_mis}
    L_{B_i,Y} = \ln\frac{\hat{p}_{B_{i}|Y}(b_{i}=1|y)}{\hat{p}_{B_{i}|Y}(b_{i}=0|y)}.
\end{equation}

 Decoding with mismatched metric $q_{B_{i},Y}(b_{i}(x),y)$ instead of $p_{Y|B_{i}}(y|b_{i})$ causes performance degradation in terms of achievable data rates. By \eqref{mis_prob_LLR} it has been shown that the maximum achievable rate is the generalized mutual information (GMI) defined as \cite{TIT09_Martinez_Fabregas_Caire_Willems} \cite{TCOM11_Nguyen_Lampe}

\begin{equation}\label{GMI_def}
    I^{GMI}_{qX,Y}=\max\limits_{s>0}I_{qX,Y}(s)
\end{equation}
with
\begin{equation}\label{I_qXY_s_B}
\begin{split}
    &I_{qX,Y}(s)\\
    &=-\sum\limits^{m-1}_{i=0}E_{X,Y}\left\{\log_{2}\sum\limits_{b\in\mathcal{B}}p_{B_{i}}(b)\left(\frac{q_{B_{i},Y}(b,Y)}{q_{B_{i},Y}(b_{i}(X),Y)}\right)^{s}\right\}\\
    &=\sum\limits^{m-1}_{i=0}\left(1-E_{X,Y}\left\{\log_{2}\left(1+\exp(-\sgn(b_i(X))L_{B_i,Y}s)\right)\right\}\right)\\
    &=\sum\limits^{m-1}_{i=0}I_{q_{B_{i},Y}}(s)
\end{split}
\end{equation}
where $L_{B_i,Y}$ is the detector output LLR that will be input to the channel decoder. 

The above shows that the GMI is the sum of $I_{q_{B_{i},Y}}(s)$ of all bit channels. $I_{q_{B_{i},Y}}(s)$ is defined as \emph{I-curve} of the $i$-th bit channel with regard to variable $s$, while $s$ that maximizes $I_{q_{B_{i},Y}}(s)$ is called the \emph{critical point}. Furthermore, it is shown that  \cite{TCOM11_Nguyen_Lampe}
\begin{equation}\label{Inequality_GMI_CbitChan_Cbicm}
    I^{GMI}_{q_{X,Y}}\leq\sum\limits^{m-1}_{i=0}I^{GMI}_{q_{B_{i},Y}}\leq\sum\limits^{m-1}_{i=0}I(B_{i},L_{i})\leq C^{BICM}.
\end{equation}


\subsection{Iterative decoding and critical point}\label{subsec:LM_MLM_issue}

It is well known that iterative decoding with the LogMAP algorithm for turbo codes (or the sum-product algorithm for LDPC codes) is approximately a maximum likelihood (ML) decoding solution when channel LLRs are optimized. However, while a non-iterative ML decoder does not need to restrict the critical point's value, iterative LogMAP decoding requires the critical point of the total I-curve to be one in order to perform well. On the other hand, iterative Max-LogMAP decoding will not be affected by changing the critical point. A shift of the total I-curve's critical point therefore has an effect equivalent to a fixed SNR mismatch. For best performance by Max-LogMAP for turbo codes (or min-sum algorithm for LDPC codes), it is sufficient to align critical points of bit channel I-curves so that they become \emph{harmonic}\footnote{Two I-curves sharing the same critical point is called harmonic \cite{TCOM11_Nguyen_Lampe}.}. For LogMAP decoding, however, we need to not only align critical points to be the same, but also align them at one. Note that if optimal MAP detection is applied, it will produce harmonic I-curves with all critical points achieved at one.

\section{Off-line based scaling functions for LLR optimization}\label{sec:off_line}

In recent literature, LLR scaling functions for GMI maximization based on off-line analysis are presented \cite{ITW10_Jalden_Fertl_Matz} \cite{TCOM11_Nguyen_Lampe}. Below we review the processes of finding scaling factors followed by a discussion on the connection between the GMI and the consistency condition.

\subsection{Optimal scaling based on off-line histogram analysis}\label{subsec:optimal_scaling_offline}
By knowledge of the given transmission scenario, two conditional PDFs can be obtained by off-line histogram estimator for LLRs from each bit channel. Denote $p_{L}(l|b=1)$ and $p_{L}(l|b=0)$ as conditional PDFs of generating LLR $l$ when bit $b=1$ and $b=0$ are transmitted, respectively. The optimal scaling factor for $l$ is
\begin{equation}\label{eq:opt_sf}
   s(l)=\frac{\ln\frac{p_{L}(l|b=1)}{p_{L}(l|b=0)}}{l}.
\end{equation}
For simple implementation, a LUT may be generated so that $s(l)$ is approximated by a piece-wise linear function and the two coefficients of each linear segment are stored for a certain range of $l$ values. In summary, this method is carried out in the following steps for each bit channel:

\begin{enumerate}
  \item Collect a sufficient amount of data of transmitted bits and detector generated LLRs of $i$-th bit channel.
  \item Calculate $p_{L}(l|b=1)$ and $p_{L}(l|b=0)$ by histograms.
  \item Obtain ideal scaling function $s_i(l)=\frac{\ln\frac{p_{L}(l|b=1)}{p_{L}(l|b=0)}}{l}$.
  \item Approximate $s_i(l)$ by piece-wise linear function $\hat{s}_i(l)$ and store linear function coefficients in a LUT.
  \item In real receiver processing, after detector generates channel LLRs $l$'s, apply a linear scaling function on each $l$ according the LUT, then continue with decoding.
\end{enumerate}

\subsection{Uniform LLR scaling based on off-line GMI maximization}\label{subsec:off_line_uniform}
 Uniform scaling of LLRs of each bit channel corresponds to a shifting of the I-curve's critical point without changing the peak value. By aligning I-curves from different bit channels to be harmonic, total GMI of BICM receiver may be increased. The following steps summarize the process of how to apply linear scaling:

\begin{enumerate}
  \item Collect a sufficient amount of data of transmitted bits and detector generated LLRs of each bit channel.
  \item Calculate and plot the I-curve $I_{q_{B_{i},Y}}(s)$ as given in \eqref{I_qXY_s_B}.
  \item Obtain a scaling factor $s_{i}$ that maximizes  $I_{q_{B_{i},Y}}(s)$.
  \item In real receiver processing, after detector generates channel LLRs $l$'s, apply scaling factor $s_i$ on each $l$ generated from $i$-th bit channel, then continue with decoding.
\end{enumerate}

\begin{figure}[h]
\centering
\includegraphics[width=3.5in]{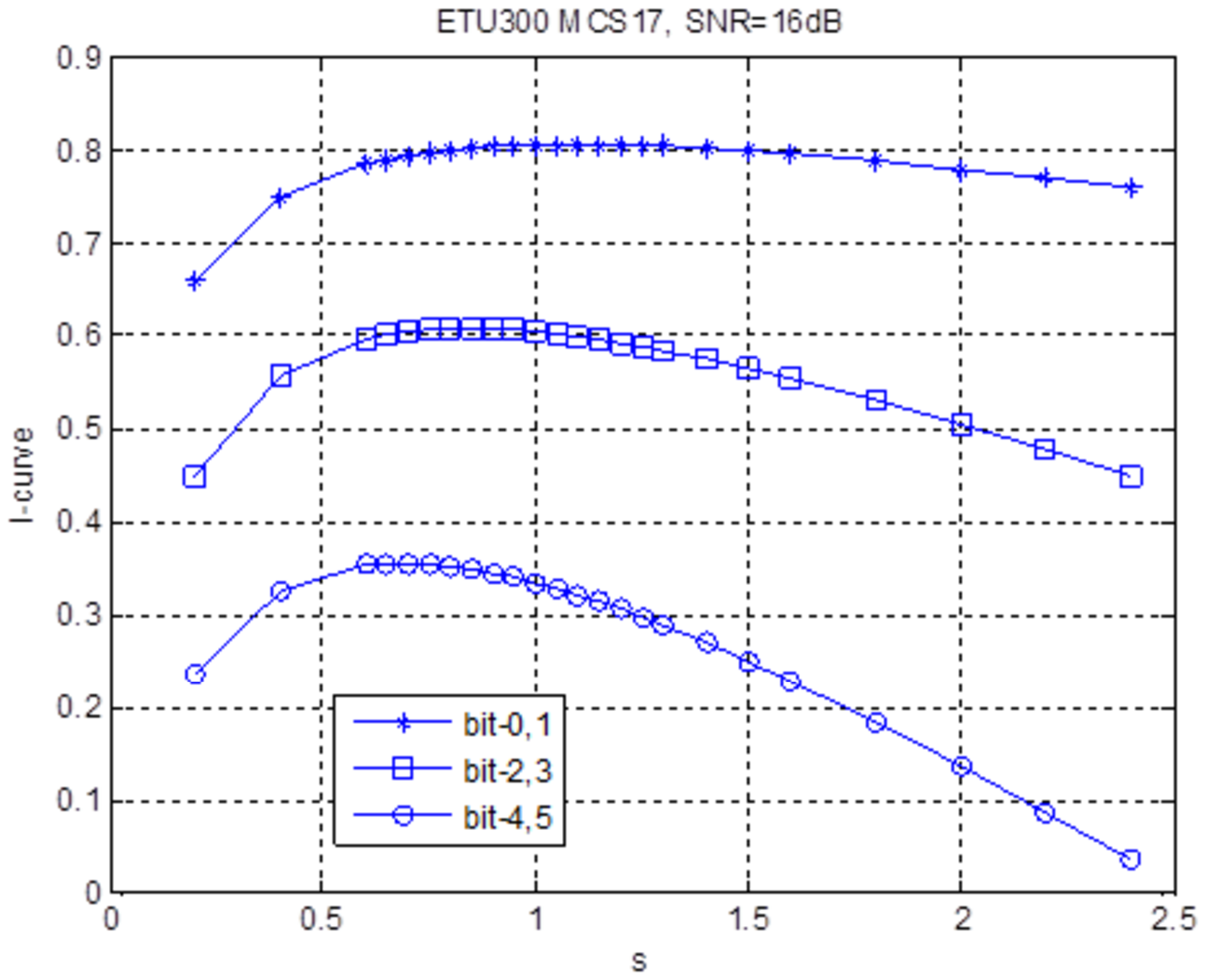}
\includegraphics[width=3.5in]{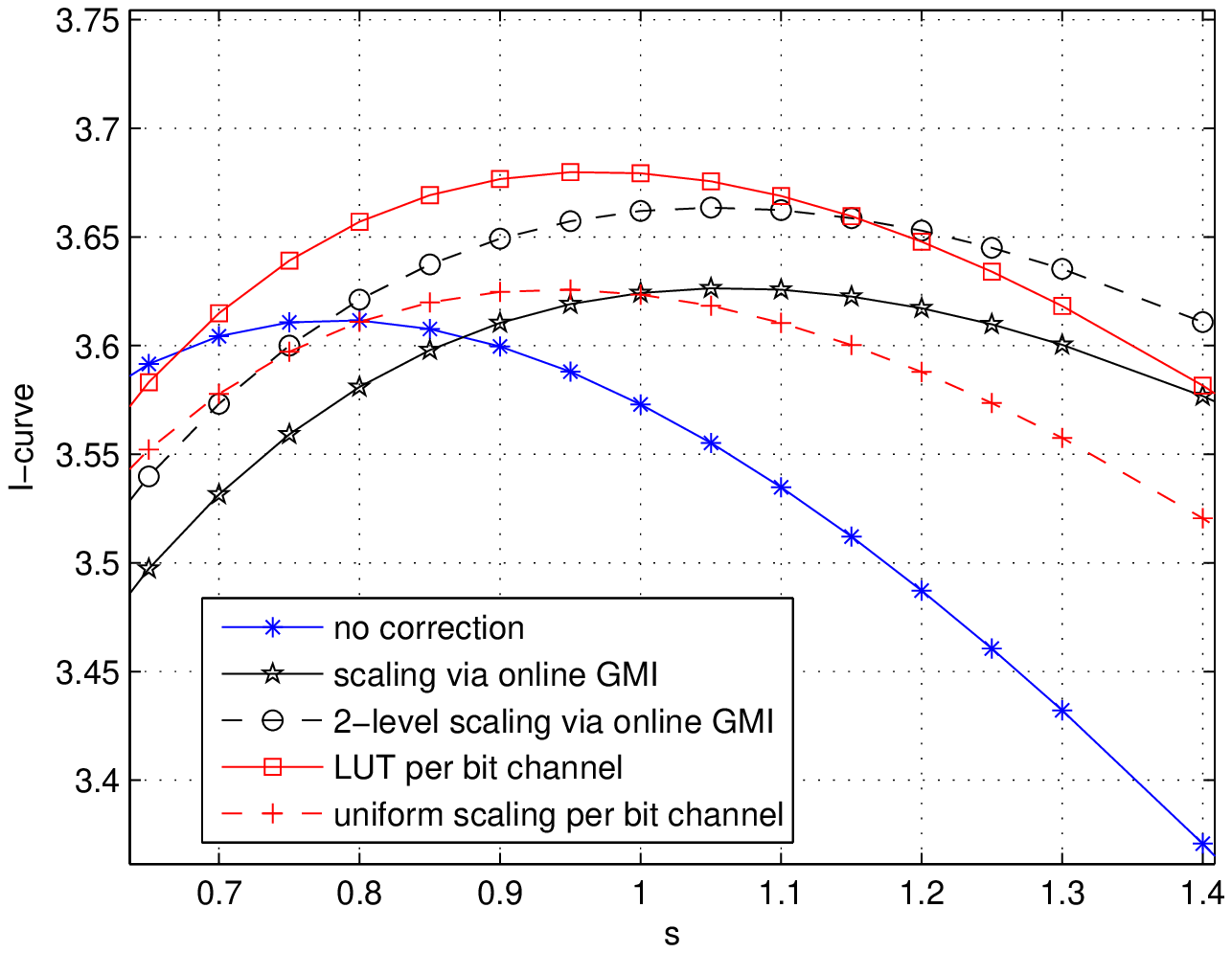}
\caption{Top: Bit channel I-curves without LLR scaling. Bottom: total I-curves before and after scaled by scalar functions found off-line and online}
\label{fig:Icurves_all}
\end{figure}

For illustration, I-curves of each bit channels obtained from link level simulation of LTE user equipment (UE) receiver are plotted in upper part of Fig.~\ref{fig:Icurves_all}. The simulated channel profile is extended typical urban (ETU) channel. Channel model parameters include excess tap delay $[0, 50, 120, 200, 230, 500, 1600, 2300, 5000]$ns with relative power $[-1.0, -1.0, -1.0, 0.0, 0.0, 0.0, -3.0, -5.0, -7.0]$dB, and Doppler frequency $300$Hz \cite{3G36101}. Transmission scheme is MCS$17$ ($64$-QAM with code rate ~$0.4$). As can be seen, critical points of each bit channel's I-curve do not align at the same value, linear scaling of LLRs can therefore increase total GMI by aligning $3$ I-curves at the same $s$. For comparison, I-curves of the total BICM channel generated from detector output LLRs without any processing, LLRs after above mentioned LUT based scaling and uniform scaling, are plotted in lower part of Fig.~\ref{fig:Icurves_all}. Clearly, LUT based scaling increases GMI more than uniform scaling does. However, both scaling methods result in the critical point achieved at $s=1$.



\subsection{Consistency condition and GMI maximization}

Below we discuss the connection between GMI maximization and the consistency condition. The consistency condition can be defined for the random variable $L$ as \cite{Hagenauer_04_EuSIPCO} \cite{icc09_Alvarado_Nunez_Szczecinski_Agrell}

\begin{equation}\label{eq_consistency}
    \ln\frac{p_{L}(l|b=1)}{p_{L}(l|b=0)}=l.
\end{equation}

While $L_{opt}$ (i.e., LLR generated from a true bit channel transition probability) satisfies \eqref{eq_consistency}, $L$ generated from \eqref{LLR_mis} generally does not. For example, in turbo decoding, consider BCJR decoding of a constituent convolutional code. As can be proven, optimal demodulation/detection generates output LLRs that satisfy consistency condition, which in turn will allow the BCJR decoder to achieve maximum a posteriori (MAP) decoding for the constituent code. In comparison, suboptimal demodulation/detection introduces two types of imperfection. $Type$-$I$: it produces less reliable estimates of transmitted coded bits, which can be modeled as a worse bit channel; $Type$-$II$ it may violate the consistency condition, which can further degrade decoding performance due to the LLR's inaccurate representation of bit probabilities. Correction of mismatched LLRs does not improve the bit channel, but can correct/reduce the inaccuracy of bit probabilities contained in the LLR values. In other words, potential performance improvement is from reducing negative effects of $Type$-$II$ but not of $Type$-$I$.

Regarding optimal scaling function that maximizes GMI (i.e., $s(l)=\frac{\ln\frac{p_{L}(l|b=1)}{p_{L}(l|b=0)}}{l}$, as shown in \cite{ITW10_Jalden_Fertl_Matz} \cite{TCOM11_Nguyen_Lampe}), we see it is equivalent to process the LLRs such that the scaled LLRs satisfy the consistency condition. For optimal uniform scaling factors, we conjecture that the uniform scaling per bit channel, that aligns each bit channels to be harmonic at $s=1$, is equal to the mean value of the optimal scaling function as
\begin{equation}\label{eq:EL_sl_consistency_condition}
    s'=E_{L}\{s(l)\}=E_{L}\left\{\frac{\ln\frac{p_{L}(l|b=1)}{p_{L}(l|b=0)}}{l}\right\}.
\end{equation}
In other words, when uniform scaling is considered, the mean value from scaling function satisfying the consistency condition not only maximizes GMI but also aligns I-curves of bit channels to be harmonic at $s=1$. Although proof of this conjecture seems difficult, numerically we have found it is supported. For example, we test max-LogMAP demodulation of $64-QAM$ signals over fast Rayleigh fading channel at SNR$=7$dB, uniform scaling factors found by using method as Section \ref{subsec:off_line_uniform} and by using $s'=E_{L}\{s(l)\}$ give us the results as in TABLE.~\ref{tb.GMIvsCC}.

\begin{table}
\renewcommand{\arraystretch}{1.3}
\caption{Scaling factors by GMI and averaging by consistency condition} \label{tb.GMIvsCC}
\begin{center}
\begin{tabular}{|c|c|c|c|c|}
\hline method/channel & bit-0,1 & bit-2,3 & bit-4,5 & total\\ \hline
$s'$ by GMI & $1.46$ & $1.39$ & $1.04$ & $1.33$ \\ \hline
$s'=E_{L}\{s(l)\}$ & $1.40$ & $1.45$ & $1.04$ & $1.32$ \\ \hline
\end{tabular}
\end{center}
\end{table}

\section{Decoder decision-aided online LLR scaling for GMI maximization}\label{sec:GMI_scaling}

Due to different statistical characteristics of LLRs obtained from different transmission scenarios, LUTs obtained in one case may be substantially different from the desirable ones for another case. The same problem occurs for the mean scaling factors obtained off-line. Therefore it is desirable to obtain scaling functions adaptive to the channel, transmission scheme, and other variations that the receiver may experience. Below we consider online uniform scaling based on GMI maximization. By adopting this approach, we may anticipate performance similar to previous off-line uniform scaling per bit channel based on GMI maximization but with real-time adaptation to channel conditions.

\subsection{Uniform scaling per bit channel}

To enable online calculation of the scaling factor for each transport block or code block, we can again apply decoder hard decisions to provide estimates of $b_i (X)$, $\hat{b}_i (X)$. As a result, we have an approximation of \eqref{I_qXY_s_B} for each transport block that characterizes current transmission conditions. For this purpose, we can calculate an approximate I-curve function for any specific $s$ as
\begin{equation}\label{I_hat}
\begin{split}
 &\hat{I}_{qB_i,Y}(s)\\
 &=1-E_{X,Y}\left\{\log_{2}\left(1+\exp(-\sgn(\hat{b}_i(X))L_{B_i,Y}s)\right)\right\}\\
       &\approx 1-\frac{1}{N_i}\sum\limits_{i=0}^{N_i-1}\log_{2}\left(1+\exp(-\sgn(\hat{b}_{i,n})L_{i,n}s)\right)
\end{split}
\end{equation}
where $N_i$ is the total number of bits transmitted through the $i$-th bit channel. The optimal scaling factor for the $i$-th bit channel is therefore
\begin{equation}\label{s_i}
        s_i=\arg\max_{s>0}\hat{I}_{qB_i,Y}(s)
\end{equation}

\begin{figure}
\centering
\includegraphics[width=3.5in]{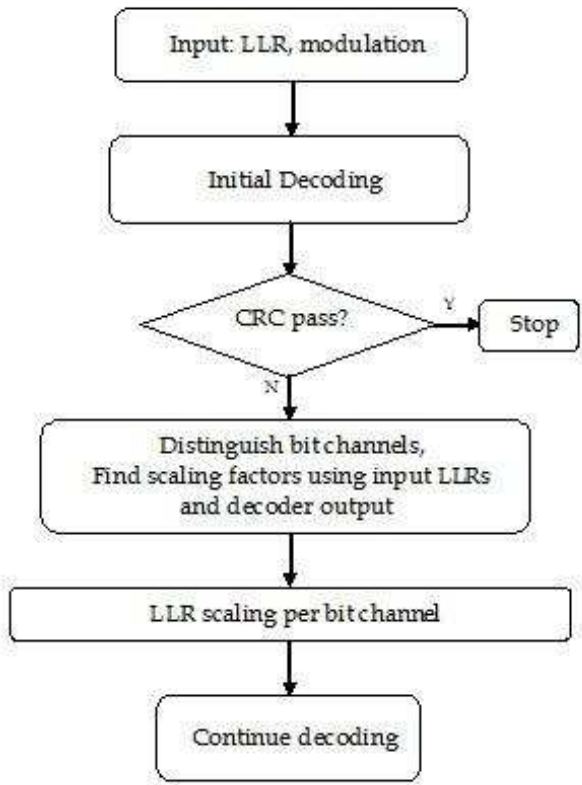}
\caption{Receiver scheme flowchart.}
\label{fig_BICM_GMI_Flowchart} \centering
\end{figure}

Based on \eqref{s_i}, the flowchart of the proposed receiver scheme after detection is shown in Fig.~\ref{fig_BICM_GMI_Flowchart}.


In practice, to identify an approximately maximized GMI, we may search for $s$ that maximizes $\hat{I}_{qB_i,Y}(s)$. Note that $\hat{I}_{qB_i,Y}(s)$ is convex with regard to $s$. To search for the optimal $s$, we can initially set $s=1$, and a multiplicative coefficient $\alpha>1$, e.g., $\alpha=1.05$, then compare $\hat{I}_{qB_i,Y}(s)$ with $\hat{I}_{qB_i,Y}(\alpha s)$ or $\hat{I}_{qB_i,Y}(s/\alpha)$ to determine whether to increase or to decrease $s$, then recursively set $s=\alpha s$ or $s=s/\alpha$ until the peak of the convex curve by $\hat{I}_{qB_i,Y}(s)$ is identified. The search process is summarized in the following $4$ steps:

Initialization: set $s=1$, $\alpha=1.05$.
\begin{enumerate}
  \item Calculate $\hat{I}_{qB_i,Y}(s)$ and $\hat{I}_{qB_i,Y}(\alpha s)$ according to \eqref{I_hat}.
  \item If $\hat{I}_{qB_i,Y}(s) > \hat{I}_{qB_i,Y}(\alpha s)$, set $\alpha=1/\alpha$. Otherwise: let  $\hat{I}_{qB_i,Y}(s)=\hat{I}_{qB_i,Y}(\alpha s)$, set  $s=\alpha s$.
  \item Calculate $\hat{I}_{qB_i,Y}(\alpha s)$. If $\hat{I}_{qB_i,Y}(s) > \hat{I}_{qB_i,Y}(\alpha s)$, set $s_i=(s+\alpha s)/2$. Stop.
  \item Otherwise, let  $\hat{I}_{qB_i,Y}(s)=\hat{I}_{qB_i,Y}(\alpha s)$, set  $s=\alpha s$. Go back to Step 3.
\end{enumerate}

For verification, we numerically calculate the I-curve obtained from the above scaling factor search algorithm for MCS17 over ETU300 with real channel and noise variance estimation. Estimation in the link level simulator is based on frequency domain minimum mean square error (MMSE) estimation using pilot symbols. It is then compared with I-curves from LLRs without scaling, off-line LUT based scaling, and off-line uniform scaling. Fig.~\ref{fig:Icurves_all} verifies that uniform scaling based on online GMI maximization can slightly increase GMI, while at the same time maximizes the I-curve approximately at $s=1$.

\subsection{Uniform scaling of positive and negative LLRs separately per bit channel}
\label{2level_per_BChan}

For $16$-QAM and $64$-QAM, due to bit loading by Gray-mapping and max-log type detection, optimal scaling functions by \eqref{eq_consistency} for LLRs from the first bit channel are symmetric around $0$, but not so for LLRs from the second and/or the third bit channels. For the second and third bit channels, it becomes reasonable to search for one scaling factor for positive LLRs and another for negative LLRs.
For verification, we plot I-curve obtained by applying different uniform scaling for positive and negative LLRs separately, and then compare with the results we obtained in Fig.~\ref{fig:Icurves_all}. We denote the new scaling method as "2-level scaling via online GMI". The results show a substantial increase in GMI is obtained by $2$-level scaling and online GMI maximization.


\subsection{Accuracy of scaling factors by decision feedback}

\begin{figure}[h]
\centering
\includegraphics[width=3.5in]{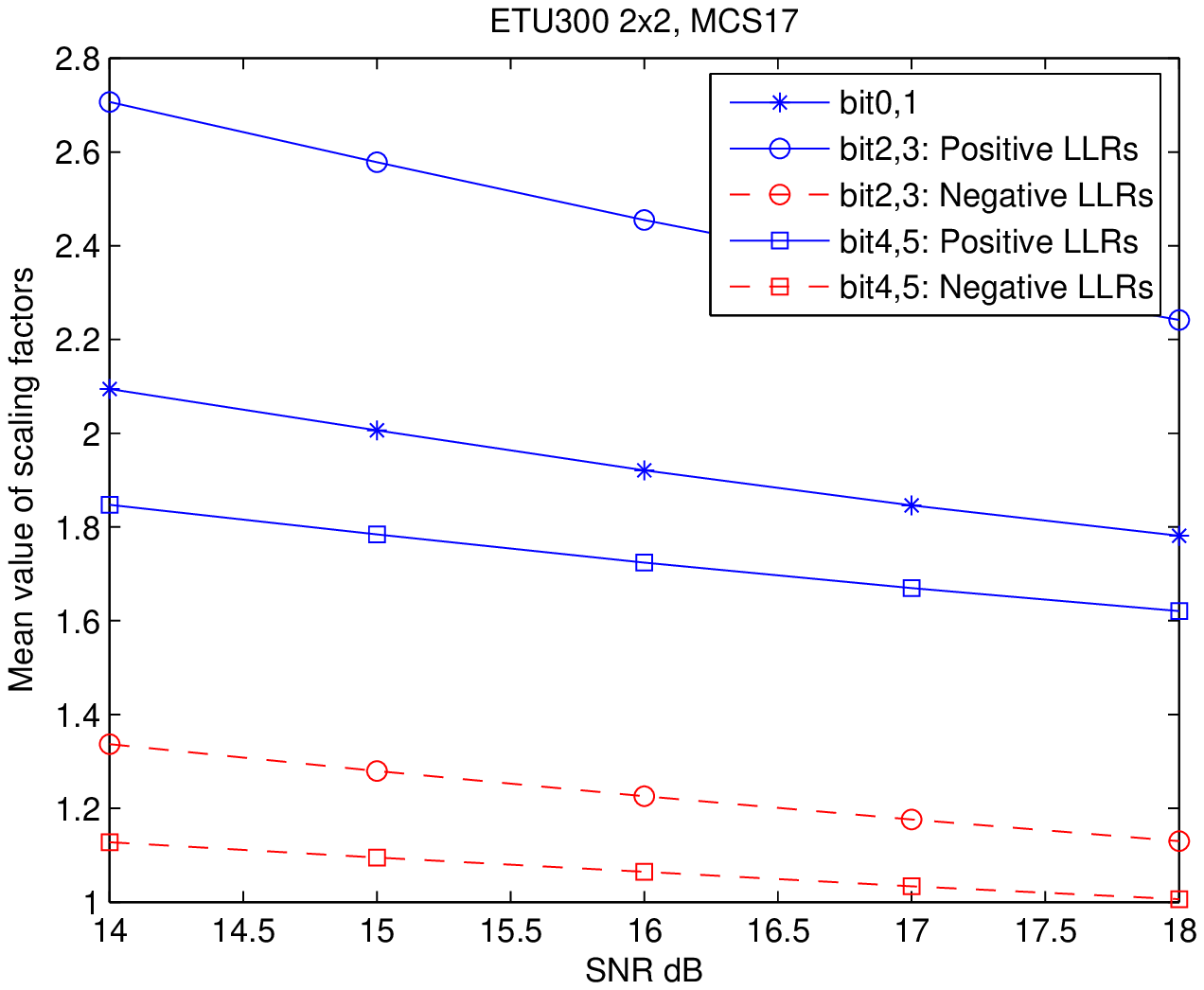}
\includegraphics[width=3.5in]{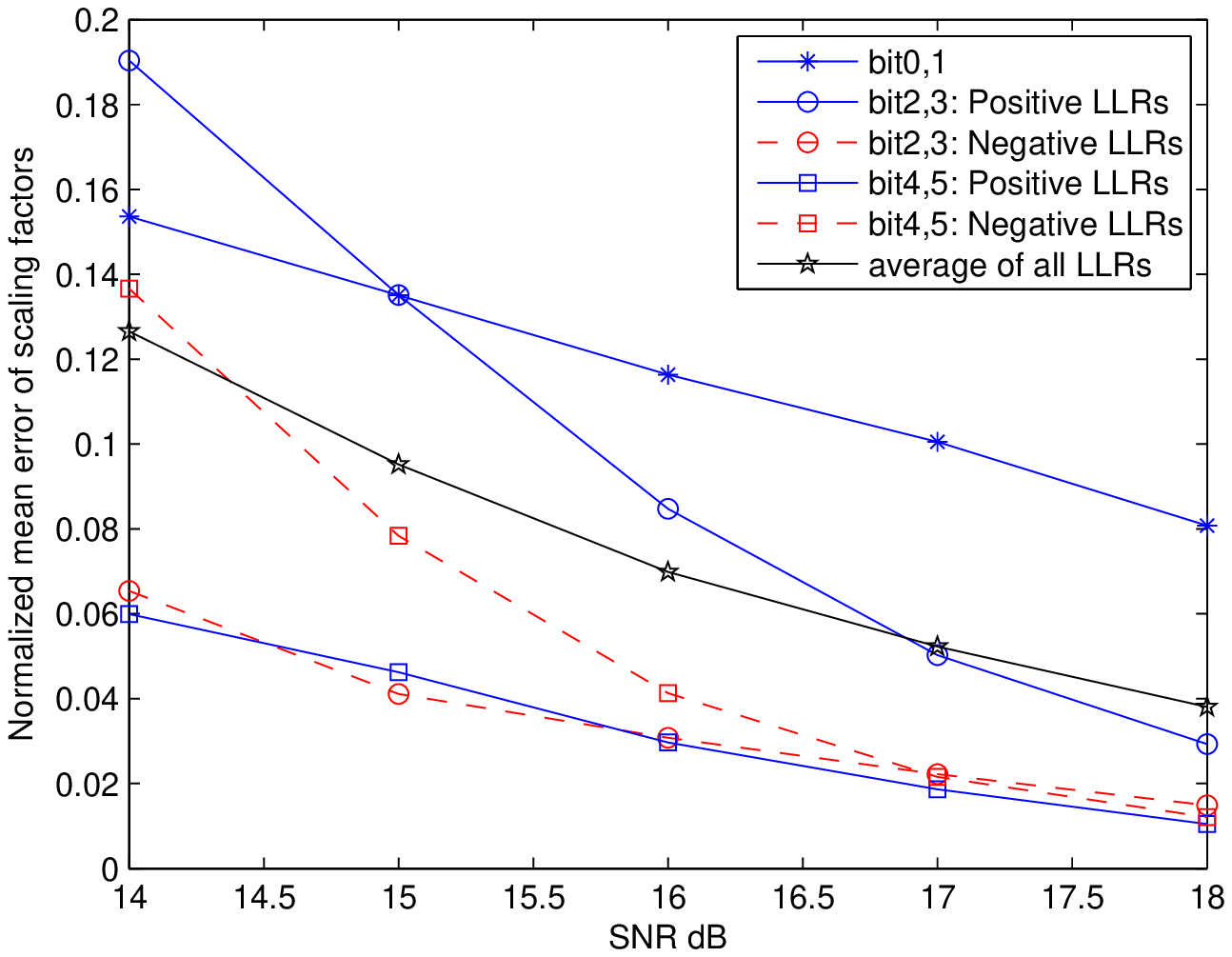}
\caption{Top: Mean scaling factors of each bit channel. Bottom: Normalized mean error of scaling factors by decoder decisions}
\label{fig:MeanErrSF_ETU300mcs17}
\end{figure}

We next study the deviation of scaling factors found by decoder decision feedback from ``genie-aided'' scaling factors that can be obtained assuming true transmitted bits are available. To quantify the extent of errors incurred by decoder decisions, we define a normalized mean error as $E\{\frac{|s-\hat{s}|}{s}\}$ where $s$ is the desired scaling factor and $\hat{s}$ is the actual available scaling factor. Numerical comparisons are carried out for the MCS$17$ over ETU$300$ channel at various SNRs. In Fig.~\ref{fig:MeanErrSF_ETU300mcs17}, the top figure shows the mean values of the ``genie'' scaling factors at different SNRs, while the bottom one plots the normalized mean error of the actual scaling factors. In these figures there is only one curve for bit channel $0$,$1$ because by Gray-mapping this channel's transition probability is symmetric for positive and negative LLRs. We make the following remarks based on Fig.~\ref{fig:MeanErrSF_ETU300mcs17}. Firstly, the mean value of the scaling factor of each bit channel gradually decreases and approaches one. This is consistent with the fact that at very high SNR, channel estimation becomes more reliable while the output by sub-optimal MLM detection also approaches that by optimal MAP detection. Secondly, overall the reliability of scaling factors from online decisions become more reliable as SNR increases, while the convergence rate toward ``genie'' decision-aided results varies by each bit channel and the sign of the LLRs.


\section{Simulations}\label{sec:simu}
\begin{figure}
\centering
\includegraphics[width=3.5in]{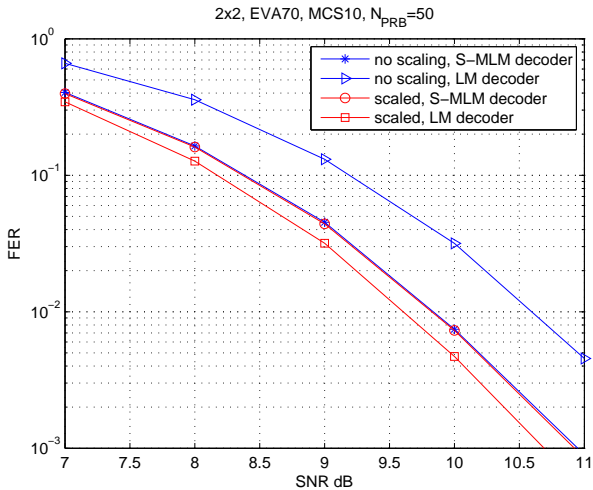}
\caption{FER: EVA70 channel, MCS10, number of physical resource block = 50, Bandwidth 10Mhz}
\label{fig_FER_EVA70mcs10} \centering
\end{figure}

\begin{figure}
\centering
\includegraphics[width=3.5in]{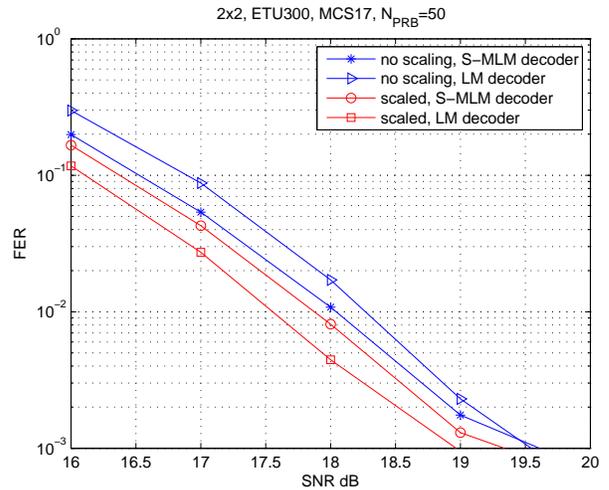}
\caption{FER: ETU300 channel, MCS17, number of physical resource block = 50, Bandwidth 10Mhz}
\label{fig_FER_ETU300mcs17} \centering
\end{figure}

We test the performance of the proposed scaling method under realistic channel conditions for an LTE system. Simulations are based on the link level simulator of LTE downlink $2\times2$ MIMO channel with open loop spatial multiplexing transmission. The system occupies $10$ MHz bandwidth using frequency division duplexing (FDD). Tested channel profiles include Extended Vehicular A (EVA) and Extended Typical Urban (ETU) channels  \cite{3G36101}. Modulation and coding sets (MCS) include MCS$10$ and MCS$17$, which correspond to $16$-QAM and $64$-QAM modulation and code rates around $0.3$ and $0.4$, respectively. At the receiver, a max-log type detector is applied. The channel decoder can be either LogMAP (LM) or scaled max-LogMAP (S-MLM) decoder. The maximum number of iterations performed by turbo decoding is $8$. For the S-MLM decoder, extrinsic LLRs of each constituent decoder are scaled by $0.7$ before being used as a priori LLR by the other constituent decoder in order to improve decoding performance \cite{vogt2000improving}. As described in Fig.~\ref{fig_BICM_GMI_Flowchart}, in order to apply the proposed LLR scaling, one initial iteration of S-MLM decoding is applied. After LLR scaling, a maximum of $7$ iterations is continued with either LM or S-MLM decoding. Frame error rate (FER) comparisons include decoding with LM or S-MLM algorithms without scaling LLRs, and decoding with the proposed scaling method. Fig.~\ref{fig_FER_EVA70mcs10}-\ref{fig_FER_ETU300mcs17} show the results. Without scaling, LM decoding performs usually worse than S-MLM decoding due to LM's sensitivity to imperfect input LLRs. However, with proper scaling, while S-MLM decoding's performance improves slightly, LM decoding's performance is improved significantly. This is consistent with the observation of I-curves in Fig.~\ref{fig:Icurves_all} where LLR scaling not only increases total GMI but also aligns the critical point at one. As expected, after LLR scaling, LM decoding outperforms S-MLM decoding substantially. From above cases, LLR scaling provides more gains over the ETU$300$ channel, which indicates that the proposed method is more effective when transmission conditions are more severe.

\section{Conclusions}\label{sec:conclusion}
 As the next generation mobile communication systems are expected to provide high throughput at expanded coverage and in versatile environments, receiver adaptation to the changing environment is important for system performance. In this work, based on the relationship between the BICM achievable rates and the GMI, we introduce a simple and effective method of LLR optimization to improve receiver performance. In order to perform LLR scaling that is adaptive to various transmission conditions, an online GMI maximization based scaling factor searching algorithm is developed. Specifically, an approximate I-curve function is calculated by using decoder decisions after certain initial decoding iteration(s), followed by a numerical search of the scaling factor that aligns the critical point of each bit channel's I-curve at one. While the effectiveness of this method depends on decoder decisions' reliability, we note that in practical operating SNR regions, the decoder decisions provide reasonable accuracy in determining the scaling factors. Increases in GMI by the proposed method are confirmed by numerical analysis. Performance improvement in realistic systems is further verified in link level simulation of LTE downlink transmissions.



\bibliography{BICM_misMetrics_Refs}

\begin{thebibliography}{1}
\providecommand{\url}[1]{#1}
\csname url@samestyle\endcsname
\providecommand{\newblock}{\relax}
\providecommand{\bibinfo}[2]{#2}
\providecommand{\BIBentrySTDinterwordspacing}{\spaceskip=0pt\relax}
\providecommand{\BIBentryALTinterwordstretchfactor}{4}
\providecommand{\BIBentryALTinterwordspacing}{\spaceskip=\fontdimen2\font plus
\BIBentryALTinterwordstretchfactor\fontdimen3\font minus
  \fontdimen4\font\relax}
\providecommand{\BIBforeignlanguage}[2]{{%
\expandafter\ifx\csname l@#1\endcsname\relax
\typeout{** WARNING: IEEEtran.bst: No hyphenation pattern has been}%
\typeout{** loaded for the language `#1'. Using the pattern for}%
\typeout{** the default language instead.}%
\else
\language=\csname l@#1\endcsname
\fi
#2}}
\providecommand{\BIBdecl}{\relax}
\BIBdecl

\bibitem{3G2585}
3GPP, ``{High Speed Downlink Packet Access: Physical Layer Aspects (Release
  5)},'' TR 25.858, Tech. Rep., 2002.

\bibitem{3G36212}
------, ``Evolved universal terrestrial radio access ({E-UTRA}); multiplexing
  and channel coding ({Release 10}),'' TS 36.212, Tech. Rep., 2011.

\bibitem{TIT09_Martinez_Fabregas_Caire_Willems}
A.~Martinez, A.~G. i~Fabregas, G.~Caire, and F.~M.~J. Willems,
  ``Bit-interleaved coded modulation revisited: A mismatched decoding
  perspective,'' \emph{{IEEE} Trans. Inform. Theory}, vol.~55, pp. 2756--2765,
  Jun. 2009.

\bibitem{TCOM11_Nguyen_Lampe}
T.~T. Nguyen and L.~Lampe, ``Bit-interleaved coded modulation with mismatched
  decoding metrics,'' \emph{{IEEE} Trans. Commun.}, vol.~59, pp. 437--447, Feb.
  2011.

\bibitem{ITW10_Jalden_Fertl_Matz}
J.~Jalden, P.~Fertl, and G.~Matz, ``On the generalized mutual information of
  bicm systems with approximate demodulation,'' in \emph{Proc. IEEE Information
  Theory Workshop (ITW)}, Cairo, Jan. 2010, pp. 1--5.

\bibitem{3G36101}
3GPP, ``Evolved universal terrestrial radio access ({E-UTRA}); user equipment
  (ue) radio transmission and reception (relasase 8),'' TS 36.101, Tech. Rep.,
  2008.

\bibitem{Hagenauer_04_EuSIPCO}
J.~Hagenauer, ``The exit chart - introduction to extrinsic information transfer
  in iterative processing,'' in \emph{European Signal Processing Conference},
  Vienna, Austria, Sep. 2004, pp. 1541--1548.

\bibitem{icc09_Alvarado_Nunez_Szczecinski_Agrell}
A.~Alvarado, V.~Nunez, L.~Szczecinski, and E.~Agrell, ``Correcting suboptimal
  metrics in iterative decoders,'' in \emph{Proc. IEEE Intl. Conf. on Com.},
  Dresden, Germany, Jun. 2009, pp. 1--6.

\bibitem{vogt2000improving}
J.~Vogt and A.~Finger, ``Improving the max-log-map turbo decoder,''
  \emph{Electronics letters}, vol.~36, no.~23, pp. 1937--1939, 2000.

\end{thebibliography}
\bibliographystyle{IEEEtran}

\end{document}